\documentclass[aps,pre,showpacs,amsmath,amssymb]{revtex4}
\usepackage{graphicx}

\begin{document}

\title{$1/f$ noise from nonlinear stochastic differential equations}

\author{J. Ruseckas}

\email{julius.ruseckas@tfai.vu.lt}

\author{B. Kaulakys}

\affiliation{Institute of Theoretical Physics and Astronomy, Vilnius
University,\\ A. Go\v{s}tauto 12, LT-01108 Vilnius, Lithuania}

\date{\today{}}

\begin{abstract}
We consider a class of nonlinear stochastic differential equations, giving the
power-law behavior of the power spectral density in any desirably wide range of
frequency. Such equations were obtained starting from the point process models
of $1/f^{\beta}$ noise. In this article the power-law behavior of spectrum is
derived directly from the stochastic differential equations, without using the
point process models. The analysis reveals that the power spectrum may be
represented as a sum of the Lorentzian spectra. Such a derivation provides
additional justification of equations, expands the class of equations
generating $1/f^{\beta}$ noise, and provides further insights into the origin
of $1/f^{\beta}$ noise.
\end{abstract}

\pacs{05.40.-a, 72.70.+m, 89.75.Da}

\maketitle

\section{Introduction.}

Power-law distributions of spectra of signals, including $1/f$ noise (also
known as $1/f$ fluctuations, flicker noise and pink noise), as well as scaling
behavior in general, are ubiquitous in physics and in many other fields,
including natural phenomena, human activities, traffics in computer networks
and financial markets. This subject has been a hot research topic for many
decades (see, e.g., a bibliographic list of papers by Li \cite{Li09}, and a
short review in Scholarpedia \cite{Scholarpedia09}).

Despite the numerous models and theories proposed since its discovery more than
80 years ago \cite{Johnson25,Schottky26}, the intrinsic origin of $1/f$ noise
still remains an open question. There is no conventional picture of the
phenomenon and the mechanism leading to $1/f$ fluctuations are not often clear.
Most of the models and theories have restricted validity because of the
assumptions specific to the problem under consideration. A short categorization
of the theories and models of $1/f$ noise is presented in the introduction of
the paper \cite{Kaulakys09}.

Until recently, probably the most general and common models, theories and
explanations of $1/f$ noise have been based on some formal mathematical
description such as fractional Brownian motion, the half-integral of the white
noise, or some algorithms for generation of signals with scaled properties
\cite{Mandelbrot68,Masry91,Milotti95,Mamontov97,Ninness98,Jumarie05,Milotti05,Ostry06,Erland07} and the popular modeling of $1/f$ noise as the superposition of
independent elementary processes with the Lorentzian spectra and a proper
distribution of relaxation times, e.g., a $1/\tau_{\mathrm{relax}}$
distribution
\cite{Bernamont37,Surdin39,Ziel50,McWhorter57,Ralls84,Rogers84,Watanabe05}. The
weakness of the latter approach is that the simulation of $1/f^{\beta}$ noise
with the desirable slope $\beta$ requires finding the special distributions of
parameters of the system under consideration; at least a wide range of
relaxation time constants should be assumed in order to allow correlation with
experiments
\cite{Hooge81,Dutta81,Weissman88,Vliet91,Wong03,Kaulakys05,Kaulakys07}.

Nonlinear stochastic differential equation with linear noise and non-linear
drift, was considered in Ref.~\cite{Mamontov97}. It was found that if the
damping is decreasing with increase of the absolute value of the stochastic
variable, then the solution of such a nonlinear SDE has long correlation time.
Recently nonlinear SDEs generating signals with $1/f$ noise were obtained in
Refs.~\cite{Kaulakys04,Kaulakys06} (see also recent papers
\cite{Kaulakys09,Kaulakys09a}), starting from the point process model of $1/f$
noise
\cite{Kaulakys981,Kaulakys982,Kaulakys991,Kaulakys992,Kaulakys001,Kaulakys002,Kaulakys003,Gontis04,Kaulakys05}.

The purpose of this article is to derive the behavior of the power spectral
density directly from the SDE, without using the point process model. Such a
derivation offers additional justification of the proposed SDE and provides
further insights into the origin of $1/f$ noise.

\section{Proposed stochastic differential equations}

Starting from the point process model, proposed and analyzed in
Refs.~\cite{Kaulakys981,Kaulakys982,Kaulakys991,Kaulakys992,Kaulakys001,Kaulakys002,Kaulakys003,Kaulakys05}, the nonlinear stochastic differential equations
are derived \cite{Kaulakys04,Kaulakys06,Kaulakys09}. The general expression for
the SDE is
\begin{equation}
dx=\sigma^2\left(\eta-\frac{\nu}{2}\right)x^{2\eta-1}dt+\sigma
x^{\eta}dW\,.
\label{eq:sde1}
\end{equation}
Here $x$ is the signal, $\eta$ is the exponent of the multiplicative noise,
$\nu$ defines the behavior of stationary probability distribution, and $W$ is a
standard Wiener process.

SDE (\ref{eq:sde1}) has the simplest form of the multiplicative noise term,
$\sigma x^{\eta}dW$. Multiplicative equations with the drift coefficient
proportional to the Stratonovich drift correction for transformation from the
Stratonovich to the It\^o stochastic equation \cite{Arnold00} generate signals
with the power-law distributions \cite{Kaulakys09}. Eq.~(\ref{eq:sde1}) is of
such type and has probability distribution of the power-law form $P(x)\sim
x^{-\nu}$. Because of the divergence of the power-law distribution and the
requirement of the stationarity of the process, the SDE (\ref{eq:sde1}) should
be analyzed together with the appropriate restrictions of the diffusion in some
finite interval. For simplicity, in this article we will adopt reflective
boundary conditions at $x=x_{\mathrm{min}}$ and $x=x_{\mathrm{max}}$. However,
other forms of restrictions are possible. For example, exponential restriction
of the diffusion can be obtained by introducing additional terms in
Eq.~(\ref{eq:sde1}),
\begin{equation}
dx=\sigma^2\left(\eta-\frac{\nu}{2}+\frac{m}{2}\left(\frac{x_{\mathrm{min}}}{
x}\right)^m-\frac{m}{2}\left(\frac{x}{x_{\mathrm{max}}}\right)^m\right)x^{2\eta
-1}dt+\sigma x^{\eta}dW\,.
\end{equation}
Here $m$ is some parameter.

Equation (\ref{eq:sde1}) with the reflective boundary condition at
$x_{\mathrm{min}}$ and $x_{\mathrm{max}}$ can be rewritten in a form that does
not contain parameters $\sigma$ and $x_{\mathrm{min}}$. Introducing the scaled
stochastic variable $x\rightarrow x/x_{\mathrm{min}}$ and scaled time
$t\rightarrow\sigma^2x_{\mathrm{min}}^{2\eta-2}t$ one transforms
Eq.~(\ref{eq:sde1}) to
\begin{equation}
dx=\left(\eta-\frac{\nu}{2}\right)x^{2\eta-1}dt+x^{\eta}dW\,.
\label{eq:sde}
\end{equation}
The scaled equation (\ref{eq:sde}) has a boundary at $x=1$ and at
\begin{equation}
\xi=\frac{x_{\mathrm{max}}}{x_{\mathrm{min}}}\,.
\end{equation}
Further we will consider Eq.~(\ref{eq:sde}) only. In order to obtain
$1/f^{\beta}$ noise we require that the region of diffusion of the stochastic
variable $x$ should be large. Therefore, we assume that $\xi\gg1$.

\section{Power spectral density from the Fokker-Planck equation}

According to Wiener-Khintchine relations, the power spectral density is
\begin{equation}
S(f)=2\int_{-\infty}^{\infty}C(t)e^{i\omega
t}dt=4\int_0^{\infty}C(t)\cos(\omega t)dt\,,
\label{eq:wk}
\end{equation}
where $\omega=2\pi f$ and $C(t)$ is the autocorrelation function. For the
stationary process the autocorrelation function can be expressed as an average
over realizations of the stochastic process,
\begin{equation}
C(t)=\langle x(t')x(t'+t)\rangle\,.
\end{equation}
This average can be written as
\begin{equation}
C(t)=\int dx\int dx'\, xx'P_0(x)P_x(x',t|x,0)\,,
\label{eq:autocorr}
\end{equation}
where $P_0(x)$ is the steady-state probability distribution function and
$P_x(x',t|x,0)$ is the transition probability (the conditional probability that
at time $t$ the signal has value $x'$ with the condition that at time $t=0$ the
signal had the value $x$). The transition probability can be obtained from the
solution of the Fokker-Planck equation with the initial condition
$P_x(x',0|x,0)=\delta(x'-x)$.

Therefore, for the calculation of the power spectral density of the signal $x$
we will use the Fokker-Planck equation instead of stochastic differential
equation (\ref{eq:sde}). The Fokker-Planck equation corresponding to the It\^o
solution of Eq.~(\ref{eq:sde}) is \cite{Gardiner85,Risken89}
\begin{equation}
\frac{\partial}{\partial
t}P=-\left(\eta-\frac{\nu}{2}\right)\frac{\partial}{\partial
x}x^{2\eta-1}P+\frac{1}{2}\frac{\partial^2}{\partial
x^2}x^{2\eta}P\,.
\label{eq:fp}
\end{equation}
The steady-state solution of Eq.~(\ref{eq:fp}) has the form
\begin{equation}
P_0(x)=
\begin{cases}
\frac{\nu-1}{1-\xi^{1-\nu}}x^{-\nu}\,, &\nu\neq1\,,\\
\frac{1}{\ln\xi}x^{-1}\,, &\nu=1\,.\end{cases}
\label{eq:stationary}
\end{equation}
The boundary conditions for Eq.~(\ref{eq:fp}) can be expressed using the
probability current \cite{Risken89}
\begin{equation}
S(x,t)=\left(\eta-\frac{\nu}{2}\right)x^{2\eta-1}P-\frac{1}{2}\frac{\partial}{
\partial x}x^{2\eta}P\,.
\label{eq:prob-current}
\end{equation}
At the reflective boundaries $x_{\mathrm{min}}=1$ and $x_{\mathrm{max}}=\xi$
the probability current $S(x,t)$ should vanish, and, therefore, the boundary
conditions for Eq.~(\ref{eq:fp}) are 
\begin{equation}
S(1,t)=0\,,\qquad S(\xi,t)=0\,.
\label{eq:boundary-1}
\end{equation}

\subsection{Eigenfunction expansion}

We solve Eq.~(\ref{eq:fp}) using the method of eigenfunctions. An ansatz of the
form
\begin{equation}
P(x,t)=P_{\lambda}(x)e^{-\lambda t}
\end{equation}
leads to the equation
\begin{equation}
-\left(\eta-\frac{\nu}{2}\right)\frac{\partial}{\partial
x}x^{2\eta-1}P_{\lambda}+\frac{1}{2}\frac{\partial^2}{\partial
x^2}x^{2\eta}P_{\lambda}=-\lambda P_{\lambda}(x)\,,
\label{eq:eigen}
\end{equation}
where $P_{\lambda}(x)$ are the eigenfunctions and $\lambda\geq0$ are the
corresponding eigenvalues. The eigenfunctions $P_{\lambda}(x)$ obey the
orthonormality relation \cite{Risken89}
\begin{equation}
\int_1^{\xi}e^{\Phi(x)}P_{\lambda}(x)P_{\lambda'}(x)dx=\delta_{
\lambda,\lambda'}\,,
\label{eq:norm-1}
\end{equation}
where $\Phi(x)$ is the potential, associated with Eq.~(\ref{eq:fp}),
\begin{equation}
\Phi(x)=-\ln P_0(x)\,.
\end{equation}
It should be noted that the restriction of diffusion of the variable $x$ by
$x_{\mathrm{min}}$ and $x_{\mathrm{max}}$ ensures that the eigenvalue spectrum
is discrete. Expansion of the transition probability density in a series of the
eigenfunctions has the form \cite{Risken89}
\begin{equation}
P_x(x',t|x,0)=\sum_{\lambda}P_{\lambda}(x')e^{\Phi(x)}P_{\lambda}(x)e^{-\lambda
 t}\,.
\label{eq:trans}
\end{equation}
Substituting Eq.~(\ref{eq:trans}) into Eq.~(\ref{eq:autocorr}) we get the
autocorrelation function
\begin{equation}
C(t)=\sum_{\lambda}e^{-\lambda t}X_{\lambda}^2\,.
\label{eq:autocorr1}
\end{equation}
Here
\begin{equation}
X_{\lambda}=\int_1^{\xi}xP_{\lambda}(x)dx
\label{eq:xx}
\end{equation}
is the first moment of the stochastic variable $x$ evaluated with the
$\lambda$-th eigenfunction $P_{\lambda}(x)$. Such an expression for the
autocorrelation function has been obtained in Ref.~\cite{Schenzle79}. Using
Eqs.~(\ref{eq:wk}) and (\ref{eq:autocorr1}) we obtain the power spectral density
\begin{equation}
S(f)=4\sum_{\lambda}\frac{\lambda}{\lambda^2+\omega^2}X_{\lambda}^2\,.
\label{eq:spectr}
\end{equation}
This expression for the power spectral density resembles the models of $1/f$
noise using the sum of the Lorentzian spectra
\cite{Bernamont37,Surdin39,Pre50,Ziel50,McWhorter57,Hooge97,Kaulakys05,Kaulakys07}. Here we see that the Lorentzians can arise from the single nonlinear
stochastic differential equation. Similar expression for the spectrum has been
obtained in Ref.~\cite{Erland07} where reversible Markov chains on finite state
spaces were considered (Eq.~(34) in Ref.~\cite{Erland07} with $-\gamma_{k,m}$
playing the role of $\lambda$ ).

A pure $1/f^{\beta}$ power spectrum is physically impossible because the total
power would be infinity. It should be noted that the spectrum of signal $x$,
obeying SDE (\ref{eq:sde}), has $1/f^{\beta}$ behavior only in some
intermediate region of frequencies, $f_{\mathrm{min}}\ll f\ll
f_{\mathrm{max}}$, whereas for small frequencies $f\ll f_{\mathrm{min}}$ the
spectrum is bounded. The behavior of spectrum at frequencies
$f_{\mathrm{min}}\ll f\ll f_{\mathrm{max}}$ is connected with the behavior of
the autocorrelation function at times $1/f_{\mathrm{max}}\ll
t\ll1/f_{\mathrm{min}}$. Often $1/f^{\beta}$ noise is described by a
long-memory process, characterized by $S(f)\sim1/f^{\beta}$ as $f\rightarrow0$.
An Abelian-Tauberian theorem relating regularly varying tails shows that this
long-range dependence property is equivalent to similar behavior of
autocorrelation function $C(t)$ as $t\rightarrow\infty$ \cite{Bingham89}.
However, this behavior of the autocorrelation function is not necessary for
obtaining required form of the power spectrum in a finite interval of the
frequencies which does not include zero \cite{Theiler91,Talocia95,Caprari98}.

From Eq.~(\ref{eq:spectr}) it follows that if the terms with small $\lambda$
dominate the sum, then one obtains $1/f^2$ behavior of the spectrum for large
frequencies $f$. If the terms with $1/f^{\beta}$ (with $\beta<2$) are present
in Eq.~(\ref{eq:spectr}), then at sufficiently large frequencies those terms
will dominate over the terms with $1/f^2$ behavior. Since the terms with small
$\lambda$ lead to $1/f^2$ behavior of the spectrum, we can expect to obtain
$1/f^{\beta}$ spectum in a frequency region where the main contribution to the
sum in Eqs.~(\ref{eq:autocorr1}) and (\ref{eq:spectr}) is from the large values
of $\lambda$. Thus we need to determine the behavior of the eigenfunctions
$P_{\lambda}(x)$ for large $\lambda$. The conditions when eigenvalue $\lambda$
can be considered as large will be investigated below.

\subsection{Eigenfunctions of the Fokker-Planck equation}

For $\eta\neq1$, it is convenient to solve Eq.~(\ref{eq:eigen}) by writing the
eigenfunctions $P_{\lambda}(x)$ in the form
\begin{equation}
P_{\lambda}(x)=x^{-\nu}u_{\lambda}(x^{1-\eta})\,.
\end{equation}
The functions $u_{\lambda}(z)$ with $z=x^{1-\eta}$ obey the equation
\begin{equation}
\frac{d^2}{dz^2}u_{\lambda}(z)-(2\alpha-1)\frac{1}{z}\frac{d}{dz}u_{
\lambda}(z)=-\rho^2u_{\lambda}(z)\,,
\label{eq:fpg}
\end{equation}
where the coefficients $\alpha$ and $\rho$ are
\begin{equation}
\alpha=1+\frac{\nu-1}{2(1-\eta)}\,,\qquad\rho=\frac{\sqrt{2\lambda}}{|\eta
-1|}\,.
\end{equation}
The area of diffusion of the variable $z=x^{1-\eta}$ is restricted by the
minimum and maximum values $z_{\mathrm{min}}$ and $z_{\mathrm{max}}$,
\begin{equation}
z_{\mathrm{min}}=
\begin{cases}
\xi^{1-\eta}\,, &\eta>1\,,\\ 1\,, &\eta<1\,,
\end{cases}\qquad z_{\mathrm{max}}=
\begin{cases}
1\,, &\eta>1\,,\\\xi^{1-\eta}\,, &\eta<1\,.\end{cases}
\end{equation}
The probability current $S_{\lambda}(x)$, Eq.~(\ref{eq:prob-current}),
rewritten in terms of functions $u_{\lambda}$, is
\begin{equation}
S_{\lambda}(z)=\frac{1}{2}(\eta-1)z^{\frac{\nu-\eta}{\eta-1}}\frac{\partial}{
\partial z}u_{\lambda}(z)\,.
\end{equation}
Therefore, the boundary conditions for Eq.~(\ref{eq:fpg}), according to
Eq.~(\ref{eq:boundary-1}) are $u_{\lambda}^{\prime}(1)=0$ and
$u_{\lambda}^{\prime}(\xi^{1-\eta})=0$. Here $u_{\lambda}^{\prime}(z)$ is the
derivative of the function $u_{\lambda}(z)$. The orthonormality relation
(\ref{eq:norm-1}) yields the orthonormality relation for functions
$u_{\lambda}(z)$,
\begin{equation}
\frac{1-\xi^{1-\nu}}{(\nu-1)(1-\eta)}\int_1^{\xi^{1-\eta}}z^{\frac{\eta-\nu}{1
-\eta}}u_{\lambda}(z)u_{\lambda'}(z)dz=\delta_{\lambda,\lambda'}\,.
\label{eq:norm}
\end{equation}
The expression (\ref{eq:xx}) for the first moment $X_{\lambda}$ of the
stochastic variable $x$ evaluated with the $\lambda$-th eigenfunction becomes
\begin{equation}
X_{\lambda}=\frac{1}{1-\eta}\int_1^{\xi^{1-\eta}}z^{\frac{1+\eta-\nu}{1
-\eta}}u_{\lambda}(z)dz\,.
\label{eq:x-u}
\end{equation}

\subsection{Solution of the equation for eigenfunctions}

The solutions of equation (\ref{eq:fpg}) are \cite{Kamke77}
\begin{equation}
u_{\lambda}(z)=z^{\alpha}\left[c_1J_{\alpha}(\rho z)+c_2Y_{\alpha}(\rho
z)\right]\,,
\label{eq:u-full}
\end{equation}
where $J_{\alpha}(z)$ and $Y_{\alpha}(z)$ are the Bessel functions of the first
and second kind, respectively. The coefficients $c_1$ and $c_2$ needs to be
determined from the boundary and normalization conditions for function
$u_{\lambda}(z)$. The asymptotic expression for the function $u_{\lambda}(z)$ is
\begin{equation}
u_{\lambda}(z)\approx
c_{\lambda}z^{\alpha-\frac{1}{2}}\rho^{-\frac{1}{2}}\cos(\rho z+a)\,,\qquad\rho
z\gg1\,.
\label{eq:g-asympt}
\end{equation}
Here $a$ is a constant to be determined from the boundary conditions and
$c_{\lambda}$ is the constant to be determined from the normalization
(\ref{eq:norm}).

The behavior of the power spectral density in Eq.~(\ref{eq:spectr}) as
$1/f^{\beta}$ can be only due to terms with large $\lambda$. Therefore, we will
consider the values of $\lambda$ for which at least the product $\rho
z_{\mathrm{max}}$ is large, $\rho z_{\mathrm{max}}\gg1$. The first moment of
the variable $x$ in the expression for the autocorrelation function
(\ref{eq:autocorr1}) is expressed via integral (\ref{eq:x-u}). If the condition
$\rho z\gg1$ is satisfied for all $z$ then the function $u_{\lambda}(z)$ has
frequent oscillations in all the region of the integration, and the integral is
almost zero. Consequently, the biggest contribution to the sum in
Eq.~(\ref{eq:autocorr1}) makes the terms corresponding to those values of
$\lambda$, for which the condition $\rho z\gg1$ is not satisfied for all values
of $z$ between $z_{\mathrm{min}}$ and $z_{\mathrm{max}}$. Therefore, we will
restrict the values of $\lambda$ by the condition $\rho z_{\mathrm{min}}\ll1$.
Thus we will consider eigenvalues $\lambda$ satisfying the conditions
\begin{equation}
1/z_{\mathrm{max}}\ll\rho\ll1/z_{\mathrm{min}}\,.
\end{equation}
Explicitly, we have conditions $1\ll\rho\ll\xi^{\eta-1}$ if $\eta>1$ and
$1/\xi^{1-\eta}\ll\rho\ll1$ if $\eta<1$.

The derivative of the function $u_{\lambda}(z)$, Eq.~(\ref{eq:u-full}), is
\begin{equation}
u_{\lambda}^{\prime}(z)=\rho z^{\alpha}\left[c_1J_{\alpha-1}(\rho
z)+c_2Y_{\alpha-1}(\rho z)\right]\,.
\label{eq:u-deriv}
\end{equation}
Since we consider the case $\rho z_{\mathrm{min}}\ll1$, then, using
Eq.~(\ref{eq:u-deriv}), instead of the boundary condition
$u_{\lambda}^{\prime}(z_{\mathrm{min}})=0$ we can approximately take the
condition 
\[
\lim_{y\rightarrow0}\left(c_1J_{\alpha-1}(y)+c_2Y_{\alpha-1}(y)\right)=0\,.
\]
If $\alpha>1$ then we get $c_2=0$; if $\alpha<1$ then
$c_2=-c_1\tan(\pi\alpha)$. Using those values of the coefficient $c_2$ we
obtain the solutions of Eq.~(\ref{eq:fpg})
\begin{equation}
u_{\lambda}(z)\approx
\begin{cases}
c_{\lambda}z^{\alpha}J_{-\alpha}(\rho z)\,, &\alpha<1\,,\\
c_{\lambda}z^{\alpha}J_{\alpha}(\rho z)\,, &\alpha>1\,.\end{cases}
\label{eq:sol-2}
\end{equation}
From approximate solution (\ref{eq:sol-2}), using asymptotic expression for the
Bessel functions, we can determine the parameter $a$ in the asymptotic
expression (\ref{eq:g-asympt}). We obtain that the parameter $a$ depnends on
$\alpha$ and does not depend on $\rho$.

\subsection{Normalization}

Taking $\lambda=\lambda'$ from Eq.~(\ref{eq:norm}) we get the normalization
condition. Using Eq.~(\ref{eq:sol-2}) we have
\[
c_{\lambda}^2\frac{1-\xi^{1-\nu}}{(\nu-1)(1-\eta)}\int_1^{\xi^{1-\eta}}zJ_{
\pm\alpha}^2(\rho
z)dz=\frac{1-\xi^{1-\nu}}{(\nu-1)}\frac{c_{\lambda}^2}{|1-\eta|\rho^2}\int_{
\rho z_{\mathrm{min}}}^{\rho z_{\mathrm{max}}}yJ_{\pm\alpha}^2(y)dy\approx1\,.
\]
Taking into account the condition $\rho z_{\mathrm{min}}\ll1$ and replacing the
lower limit of integration by $0$, we obtain that the integral is approximately
equal to
\[
\int_0^{\rho z_{\mathrm{max}}}yJ_{\pm\alpha}^2(y)dy\approx\frac{\rho
z_{\mathrm{max}}}{\pi}\,.
\]
Here we assumed that $\rho z_{\mathrm{max}}\gg1$. Therefore, the normalization
constant $c_{\lambda}$ is 
\begin{equation}
c_{\lambda}\approx\sqrt{\frac{|1-\eta|}{z_{\mathrm{max}}}\frac{\nu-1}{1-\xi^{1
-\nu}}\pi\rho}\,.
\label{eq:c-lambda}
\end{equation}

\section{Calculation of the power spectral density}

\subsection{Estimation of the first moment $X_{\lambda}$ of the stochastic
variable $x$}

The expression (\ref{eq:autocorr1}) for the autocorrelation function contains
the first moment $X_{\lambda}$ of the variable $x$, expressed as an integral of
the function $u_{\lambda}$, Eq.~(\ref{eq:x-u}). Using Eq.~(\ref{eq:sol-2}) we
get
\[
X_{\lambda}\approx\frac{c_{\lambda}}{1-\eta}\int_1^{\xi^{1-\eta}}z^{\beta-1}J_{
\pm\alpha}(\rho z)dz=\frac{c_{\lambda}}{|1-\eta|\rho^{\beta}}\int_{\rho
z_{\mathrm{min}}}^{\rho z_{\mathrm{max}}}y^{\beta-1}J_{\pm\alpha}(y)dy\,.
\]
Here
\begin{equation}
\beta=1+\frac{\nu-3}{2(\eta-1)}
\end{equation}
and ``$+$'' sign is for $\alpha>1$, while ``$-$'' is for $\alpha<1$.

If $\pm\alpha+\beta>0$, taking into account that $\rho z_{\mathrm{min}}\ll1$,
we can replace the lower limit of the integration by $0$,
$X_{\lambda}\approx\frac{c_{\lambda}}{|1-\eta|}\frac{1}{\rho^{\beta}}\int_0^{
\rho z_{\mathrm{max}}}y^{\beta-1}J_{\pm\alpha}(y)dy$ . We get that the integral
in the expression for $X_{\lambda}$ approximately does not depend on the lower
limit of integration $\rho z_{\mathrm{min}}$. We can integrate the integral by
parts and use the properties of the Bessel functions to obtain
\begin{equation}
X_{\lambda}\approx\frac{c_{\lambda}}{|1-\eta|\rho^{\beta}}\left(\mp\left.y^{
\beta-1}J_{\pm(\alpha-1)}(y)\right|_{\rho z_{\mathrm{min}}}^{\rho
z_{\mathrm{max}}}\pm(\beta+\alpha-2)\int_{\rho z_{\mathrm{min}}}^{\rho
z_{\mathrm{max}}}y^{\beta-2}J_{\pm(\alpha-1)}(y)dy\right)\,.
\label{eq:i2parts}
\end{equation}
Using expression (\ref{eq:sol-2}) for the function $u_{\lambda}(z)$, the
boundary conditions $u_{\lambda}^{\prime}(z_{\mathrm{min}})=0$ and
$u_{\lambda}^{\prime}(z_{\mathrm{max}})=0$ leads to
\begin{equation}
J_{\pm(\alpha-1)}(\rho z_{\mathrm{min}})=0\,,\qquad J_{\pm(\alpha-1)}(\rho
z_{\mathrm{max}})=0\,.
\end{equation}
Therefore, the first term in the expression (\ref{eq:i2parts}) for
$X_{\lambda}$ is zero. If $\beta<\frac{5}{2}$, taking into account that $\rho
z_{\mathrm{max}}\gg1,$ we can extend the upper limit of integration to
$+\infty$. We get that the integral for $X_{\lambda}$ approximately does not
depend on the upper limit of integration $\rho z_{\mathrm{max}}$.

Therefore, the first moment $X_{\lambda}$ of the variable $x$ is proportional
to the expression
\begin{equation}
\frac{c_{\lambda}}{|1-\eta|}\frac{1}{\rho^{\beta}}\,.
\label{eq:i-2}
\end{equation}
Now we are ready to estimate the power spectral density.

\subsection{Power spectral density}

Since $\rho z_{\mathrm{max}}\gg1$ , from the boundary condition
$u_{\lambda}^{\prime}(z_{\mathrm{max}})=0$ using the asymptotic expression
(\ref{eq:g-asympt}) for the function $u_{\lambda}(z)$ we obtain the condition
$\sin(\rho z_{\mathrm{max}}+a)=0$ and $\rho z_{\mathrm{max}}=\pi n-a$. Then
\begin{equation}
\lambda_n\approx\frac{(1-\eta)^2}{2z_{\mathrm{max}}^2}(\pi
n-a)^2\,.
\label{eq:lambdan}
\end{equation}
Equation (\ref{eq:lambdan}) shows that the density of eigvalues $D(\lambda)$ is
proportional to $1/\sqrt{\lambda}$. Since the parameter $a$ does not depend on
$\lambda$, it follows that the density of eigenvalues and, consequently, the
autocorrelation function do not depend on th parameter $a$.

In order to estimate the sum in expression (\ref{eq:autocorr1}) for the
autocorrelation function, we replace summation by the integration,
\begin{equation}
C(t)\approx\int e^{-\lambda t}X_{\lambda}^2D(\lambda)d\lambda
\end{equation}
Such a replacement is valid when $\rho z_{\mathrm{max}}\gg1$. Using the
approximate expressions (\ref{eq:c-lambda}) and (\ref{eq:i-2}) we get the
expression for the autocorrelation function
\begin{equation}
C(t)\sim\int_{z_{\mathrm{max}}^{-2}}^{z_{\mathrm{min}}^{-2}}\lambda^{-\beta}e^{
-\lambda
t}d\lambda=t^{\beta-1}\left[\Gamma(1-\beta,z_{\mathrm{max}}^{-2}t)-\Gamma(1
-\beta,z_{\mathrm{min}}^{-2}t)\right]\,.
\label{eq:autocorr-approx}
\end{equation}
Here $\Gamma(a,z)=\int_z^{\infty}t^{a-1}e^{-t}dt$ is the incomplete Gamma
function. When $z_{\mathrm{min}}^2\ll t\ll z_{\mathrm{max}}^2$ we have the
following lowest powers in the expansion of the approximate expression
(\ref{eq:autocorr-approx}) for the autocorrelation function in the power series
of $t$:
\begin{equation}
C(t)\sim
\begin{cases}
\frac{z_{\mathrm{max}}^{2(\beta-1)}}{\beta-1}-\frac{tz_{\mathrm{max}}^{2(\beta
-2)}}{\beta-2}\,, &\beta>2\\
z_{\mathrm{max}}^2+(\gamma-1)t+t\ln(z_{\mathrm{max}}^{-2}t)\,, &\beta=2\\
\frac{z_{\mathrm{max}}^{2(\beta-1)}}{\beta-1}+t^{\beta-1}\Gamma(1-\beta)\,, &
1<\beta<2\\ -\gamma-\ln(z_{\mathrm{max}}^{-2}t)\,, &\beta=1\\
\frac{1}{t^{1-\beta}}\Gamma(1-\beta)\,, &\beta<1\end{cases}
\end{equation}
Here $\gamma\approx0.577216$ is the Euler's constant. Similar first terms in
the expansion of the autocorrelation function in the power series of time $t$
has been obtained in Ref.~\cite{Kaulakys09}.

Similarly, when $\rho z_{\mathrm{max}}\gg1$, replacing in Eq.~(\ref{eq:spectr})
the summation by the integration we obtain the power spectral density
\begin{equation}
S(f)\approx4\int\frac{\lambda}{\lambda^2+\omega^2}X_{
\lambda}^2D(\lambda)d\lambda\,.
\label{eq:spectr-int}
\end{equation}
Equation, similar to Eq.~(\ref{eq:spectr-int}) has been obtained in
Ref.~\cite{Milotti95} by considering a relaxing linear system driven by white
noise (Eq.~(27) in Ref.~\cite{Milotti95}). Similar equation also has been
obtained in Ref.~\cite{Erland07} where reversible Markov chains on finite state
spaces were considered. In both Refs.~\cite{Milotti95}, \cite{Erland07} the
power spectral density is expressed as a sum or an integral over the
eigenvalues of a matrix describing transitions in the system.

Using the approximate expressions (\ref{eq:c-lambda}) and (\ref{eq:i-2}) we get
the equation
\begin{equation}
S(f)\sim\int_{z_{\mathrm{max}}^{-2}}^{z_{\mathrm{min}}^{-2}}\frac{1}{\lambda^{
\beta-1}}\frac{1}{\lambda^2+\omega^2}d\lambda\,.
\label{eq:spectr-approx}
\end{equation}
When $z_{\mathrm{max}}^{-2}\ll\omega\ll z_{\mathrm{min}}^{-2}$ then the leading
term in the expansion of the approximate expression (\ref{eq:spectr-approx})
for the power spectral density in the power series of $\omega$ is
\begin{equation}
S(f)\sim
\begin{cases}
\omega^{-\beta}\,, &\beta<2\,,\\\omega^{-2}\,, &\beta\ge2\,.\end{cases}
\end{equation}
The second term in the expansion is proportional to $\omega^{-2}$. In the case
of $\beta<2$, the term with $\omega^{-\beta}$ becomes larger than the term with
$\omega^{-2}$ when $z_{\mathrm{max}}^{-2}\ll\omega$. Therefore, we obtain
$1/f^{\beta}$ spectrum in the frequency interval $1\ll\omega\ll\xi^{2(\eta-1)}$
if $\eta>1$ and the frequency interval $1/\xi^{2(1-\eta)}\ll\omega\ll1$ if
$\eta<1$. It should be noted that time $t$ and frequency $\omega$ in our
analysis are dimensionless.

Eq.~(\ref{eq:spectr-int}) shows that the shape of the power spectrum depends on
the behavior of the eigenfunctions and the eigenvalues in terms of the function
$X_{\lambda}^2D(\lambda)$. This function $X_{\lambda}^2D(\lambda)$ should be
proportional to $\lambda^{-\beta}$ in order to obtain $1/f^{\beta}$ behavior.
Similar condition has been obtained in Ref.~\cite{Erland07}. Eq.~(13) for
discrete time process and the unnumbered equation after Eq.~(34) for a
continuous time process in Ref.~\cite{Erland07} are analogous to the condition
$X_{\lambda}^2D(\lambda)\sim\lambda^{-\beta}$ since the density of eigenvalues
in Ref.~\cite{Erland07} is proportional to $1/|\gamma^{\prime}(x)|$. Equations
(\ref{eq:autocorr-approx}) and (\ref{eq:spectr-approx}) for the power spectral
density and autocorrelation function resembles those obtained from the sum of
Lorentzian signals with appropriate weights in Ref.~\cite{Kaulakys05}.

\section{Numerical examples}

If $\nu=3$ we get that $\beta=1$ and stochastic differential equation
(\ref{eq:sde1}) should give signal exhibiting $1/f$ noise. We will solve
numerically two cases: $\eta=\frac{5}{2}>1$ and $\eta=-\frac{1}{2}<1$. For the
numerical solution we use Euler-Marujama approximation, transforming
differential equations to difference equations. Equation (\ref{eq:eta-5-2})
with $\eta=5/2$ was solved using variable step of integration, solution
Eq.~(\ref{eq:eta-1-2}) with $\eta=-1/2$ was performed using a fixed step of
integration.

\begin{figure}
\begin{centering}
\includegraphics[width=0.45\textwidth]{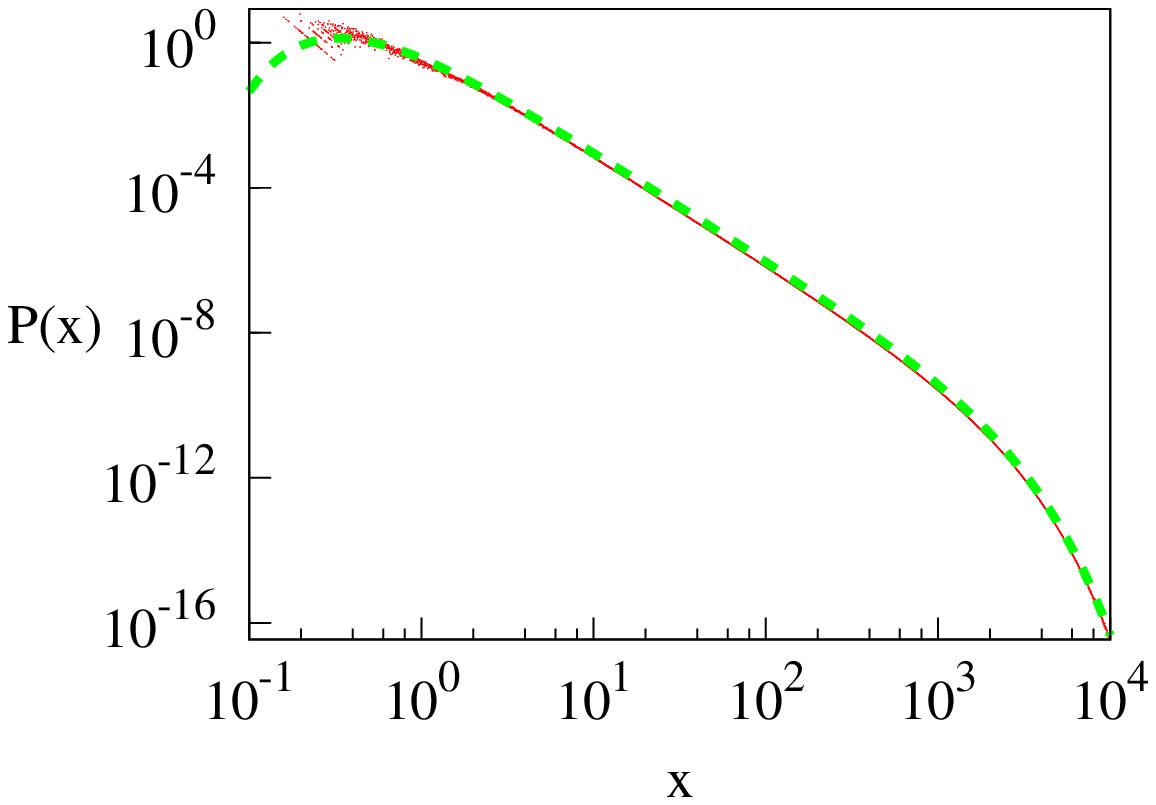}\includegraphics[width=0.45\textwidth]{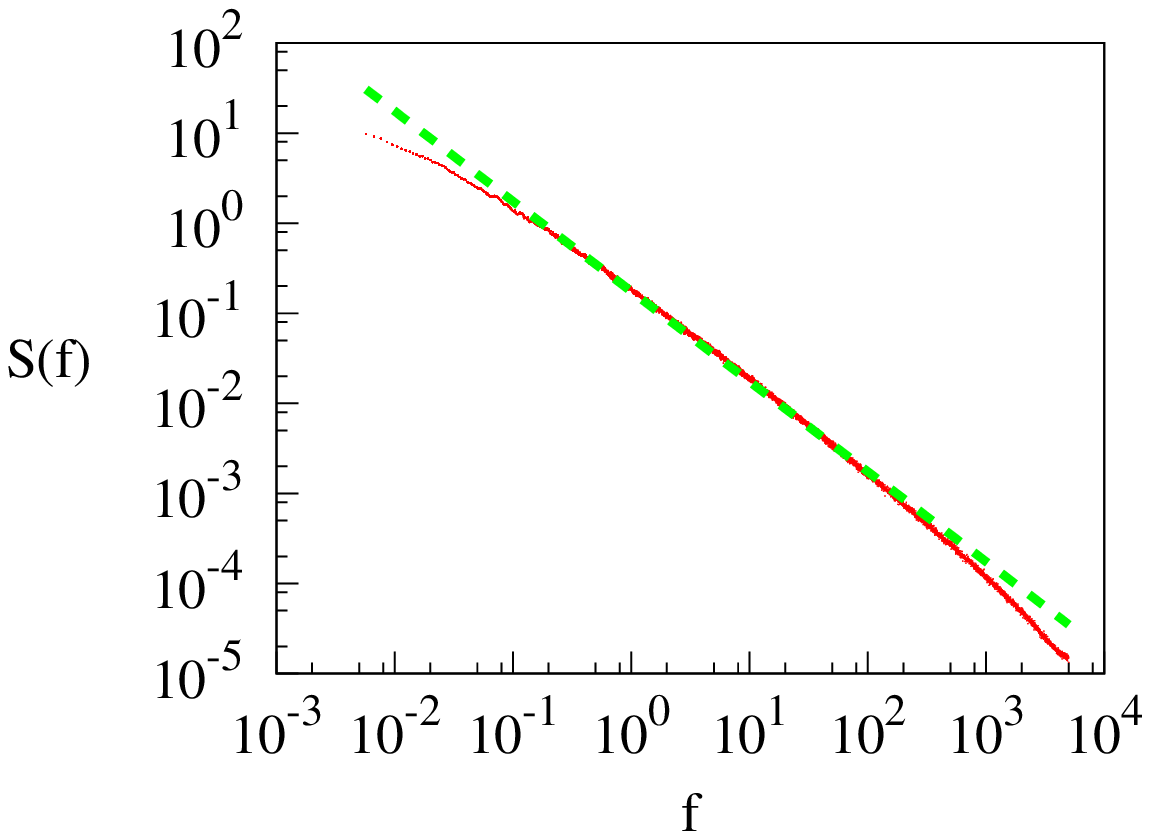}\caption{(Color online) Probability
distribution function $P(x)$ (left) and power spectral density $S(f)$ (right)
for the stochastic process defined by the stochastic differential equation
(\ref{eq:eta-5-2}). Dashed green lines are analytical expression
(\ref{eq:stationary}) for the steady-state distribution function $P_0(x)$ on
the left and the slope $1/f$ on the right. Parameters used are
$x_{\mathrm{min}}=1$, $x_{\mathrm{max}}=10^2$, and $\sigma=1$ .}
\label{fig:eta-5-2} \par
\end{centering}

\end{figure}

When $\eta=\frac{5}{2}$ and $\nu=3$ then equation (\ref{eq:sde1}) is
$dx=\sigma^2x^4dt+\sigma x^{\frac{5}{2}}dW$ . Using exponential restriction of
the diffusion region we have the equation 
\begin{equation}
dx=\sigma^2\left(1+\frac{1}{2}\frac{x_{\mathrm{min}}}{x}-\frac{1}{2}\frac{x}{x_{
\mathrm{max}}}\right)x^4dt+\sigma x^{\frac{5}{2}}dW\,.
\label{eq:eta-5-2}
\end{equation}
The equation was solved using the variable step of integration, $\Delta
t_k=\kappa^2/x_k^3$, with $\kappa\ll1$ being a small parameter. The
steady-state probability distribution function $P_0(x)$ and the power spectral
density $S(f)$ are presented in Fig.~\ref{fig:eta-5-2}. We see a good agreement
of the numerical results with the analytical expressions. The $1/f$ interval in
the power spectral density in Fig.~\ref{fig:eta-5-2} is approximately between
$f_{\mathrm{min}}\approx2\times10^{-1}$ and
$f_{\mathrm{max}}\approx2\times10^2$. The width of this region is much narrower
than the width of the region $1\ll2\pi f\ll10^6$ ($\xi=10^2$) predicted in the
previous section.

\begin{figure}

\begin{centering}
\includegraphics[width=0.45\textwidth]{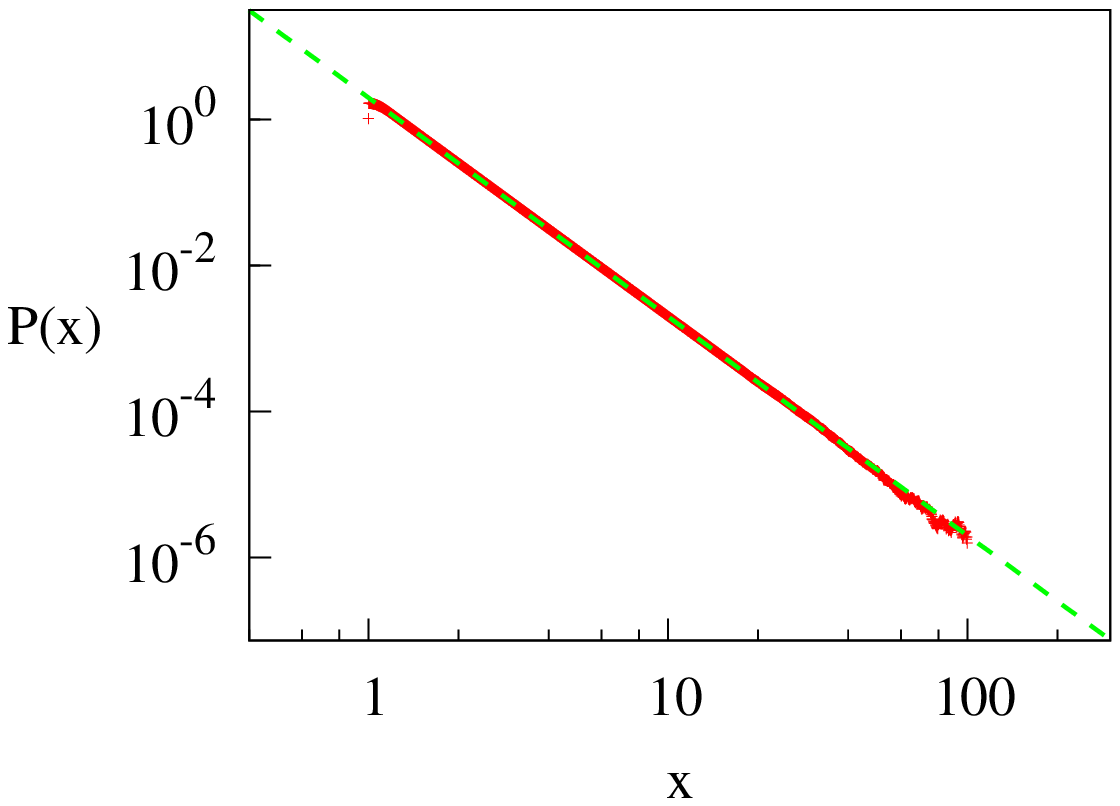}\includegraphics[width=0.45\textwidth]{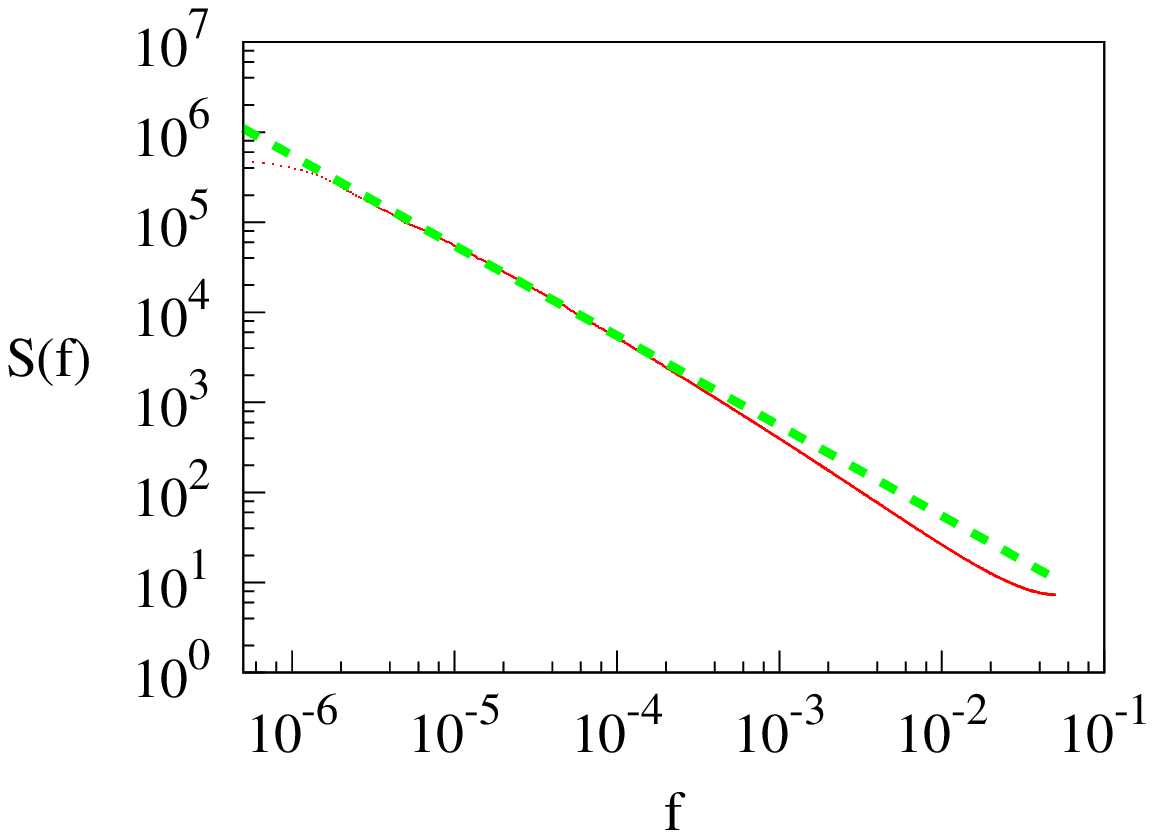}\caption{(Color online) Probability
distribution function $P(x)$ (left) and power spectral density $S(f)$ (right)
for the stochastic process defined by the stochastic differential equation
(\ref{eq:eta-1-2}). Dashed green lines are analytical expression
(\ref{eq:stationary}) for the steady-state distribution function $P_0(x)$ on
the left and the slope $1/f$ on the right. Parameters used are
$x_{\mathrm{min}}=1$, $x_{\mathrm{max}}=10^2$, and $\sigma=1$ .}
\label{fig:eta-1-2} \par
\end{centering}

\end{figure}

When $\eta=-1/2$ and $\nu=3$ then equation (\ref{eq:sde1}) is
\begin{equation}
dx=-2\frac{\sigma^2}{x^2}dt+\frac{\sigma}{\sqrt{x}}dW
\label{eq:eta-1-2}
\end{equation}
We used reflective boundary conditions at $x_{\mathrm{min}}=1$ and
$x_{\mathrm{max}}=100$. The equation was solved with a constant step of
integration. The steady-state probability distribution function $P_0(x)$ and
the power spectral density $S(f)$ are presented in Fig.~\ref{fig:eta-1-2}. The
$1/f$ interval in the power spectral density in Fig.~\ref{fig:eta-1-2} is
approximately between $f_{\mathrm{min}}\approx10^{-6}$ and
$f_{\mathrm{max}}\approx2\times10^{-4}$. The width of this region is much
narrower than the width of the region $10^{-6}\ll2\pi f\ll1$ ($\xi=10^2$)
predicted in the previous section.

Numerical solution of the equations confirms the presence of the frequency
region for which the power spectral density has $1/f^{\beta}$ dependence. The
width of this region can be increased by increasing the ratio between minimum
and maximum values of the stochastic variable $x$. In addition, the region in
the power spectral density with the power-law behavior depends on the exponent
$\eta$: if $\eta=1$ then this width is zero; the width increases with
increasing the difference $|\eta-1|$. However, the estimation of the width of
the region, obtained in the previous section, is too broad, the width obtained
in numerical solutions is narrower. Such a discrepancy can be explained as the
result of various approximations, made in the derivation.

\section{Discussion}

In summary, we derived the behavior of the power spectral density from the
nonlinear stochastic differential equation. In
Refs.~\cite{Kaulakys04,Kaulakys06} only the values of the exponent of the
multiplicative noise $\eta$ greater than $1$ has been used. Here we showed,
that it is possible to obtain $1/f^{\beta}$ noise from the same nonlinear SDE
for $\eta<1$, as well. The analysis reveals that the power spectrum may be
represented as a sum of the Lorentzian spectra with the coefficients
proportional to the squared first moments of the stochastic variable evaluated
with the appropriate eigenfunctions of the corresponding Fokker-Planck
equation. Nonlinear SDE, corresponding to a particular case of
Eq.~(\ref{eq:sde1}) with $\eta=0$, i.e., with linear noise and non-linear
drift, was considered in Ref.~\cite{Mamontov97}. It was found that if the
damping is decreasing with increase of $|x|$, then the solution of such a
nonlinear SDE has long correlation time.

As Eq.~(\ref{eq:spectr-int}) shows, the shape of the power spectrum depends on
the behavior of the eigenfunctions and the eigenvalues in terms of the function
$X_{\lambda}^2D(\lambda)$, where $D(\lambda)$ is the density of eigenvalues.
The SDE (\ref{eq:sde}) considered in this article gives the density of
eigenvalues $D(\lambda)$ proportional to $1/\sqrt{\lambda}$. One obtains
$1/f^{\beta}$ behavior of the power spectrum when this function
$X_{\lambda}^2D(\lambda)$ is proportional to $\lambda^{-\beta}$ for a wide
range of eigenvalues $\lambda$, as is the case for SDE (\ref{eq:sde}). Similar
condition has been obtained in Ref.~\cite{Erland07}.

One of the reasons for the appearance of the $1/f^{\beta}$ spectrum is the
scaling property of the stochastic differential equation (\ref{eq:sde1}):
changing the stochastic variable from $x$ to $x'=ax$ changes the time-scale of
the equation to $t'=a^{2(1-\eta)}t$ , leaving the form of the equation
invariant. From this property it follows that it is possible to eliminate the
eigenvalue $\lambda$ in Eq.~(\ref{eq:eigen}) by changing the variable from $x$
to $z=\lambda^{1/2(1-\eta)}x$. The dependence of the eigenfunction on
eigenvalue $\lambda$ then enters only via the boundary conditions. Such scaling
properties were used estimating the norm of the eigenfunction and the first
moment $X_{\lambda}$ of the stochastic variable $x$ evaluated with the
$\lambda$-th eigenfunction. Other factor in obtaining the power-law spectrum is
wide range of the region of diffusion of the stochastic variable $x$.

\end{document}